\documentclass{aa}  

\usepackage{graphicx}
\usepackage{txfonts}
\usepackage{hyperref}
\bibpunct{(}{)}{;}{a}{}{,} 

\newcommand{\spics}{_small}

\newcommand{\fdir}{./}
\newcommand{\oebdir}{./}

\newcommand{\xmm}{{\it XMM-Newton~}}
\newcommand{\chandra}{{\it Chandra~}}
\newcommand{\rosat}{{\it ROSAT~}}
\newcommand{\fermi}{{\it FERMI~}}

\newcommand{\nbr}[1]{\left( #1 \right)} 
\newcommand{\sbr}[1]{\left[ #1 \right]} 

\newcommand{\rv}[1]{{ #1 }}

\begin{document}

 \title{Feedback by massive stars and the emergence of superbubbles}
\titlerunning{Emergence of superbubbles -- X-ray properties}
\subtitle{II. X-ray properties}
 \author{Martin Krause \inst{1,2} \fnmsep\thanks{E-mail:
    krause@mpe.mpg.de} \and Roland Diehl \inst{1,2}
  \and Hans B\"ohringer \inst{1,2} \and Michael Freyberg \inst{1}
\and Daniel Lubos \inst{1}}

\institute{
  Max-Planck-Institut f\"ur extraterrestrische Physik,
  Giessenbachstr.~1, 85741 Garching, Germany \and
Excellence Cluster Universe, Technische Universit\"at
  M\"unchen, Boltzmannstrasse 2, 85748 Garching, Germany 
}

   \date{Received March 24, 2014; accepted ?}

 
  \abstract
   {In a previous paper we investigated the energy transfer of
     massive stars to the interstellar medium as a function of time
     and the geometrical configuration of three massive stars via 
     3D-mesh-refining hydrodynamics simulations, following the
     complete evolution of the massive stars and their supernovae
	\rv{except non-thermal processes}.}
   {In order to compare our results against observations we
     derive \rv{thermal} X-ray properties of the interstellar medium 
	from our simulations and compare them to
     observations of superbubbles in general, to the
     well-studied nearby \object{Orion-Eridanus} superbubble and to the 
     diffuse soft X-ray emission of nearby galaxies. }
   {We analysed our ISM simulation results with the help of spectra for plasma
     temperatures between 0.1 and  10 keV and computed the spectral evolution and
     the spatio-temporal distribution of the hot gas.}
   {Despite significant input of high temperature gas from supernovae
     and fast stellar winds, the resulting \rv{thermal} X-ray spectra are generally very soft, with
   most of the emission well below 1 keV. We show that this is
   due to mixing triggered by resolved hydrodynamic
   instabilities. Supernovae enhance the X-ray luminosity of a
   superbubble by 1-2 orders of magnitude for a time span of about
   0.1~Myr; longer if a supernova occurs in a larger superbubble and
   shorter in higher energy bands. Peak superbubble luminosities of the
   order of $10^{36}$~erg~s$^{-1}$ are reproduced well.
   The strong decay of the X-ray luminosity is due to bubble expansion, 
   hydrodynamic instabilities related to the
   acceleration of the superbubble's shell thanks to the sudden energy
   input, and subsequent mixing.  We also find global oscillations of
   our simulated superbubbles, which produce spatial variations of the
   X-ray spectrum, similar to what we see in the Orion-Eridanus
   cavity. We calculated the fraction of energy emitted in X-rays and
   find that with a value of a few times $10^{-4}$, it is about a factor of ten below the measurements for
   nearby galaxies.
 }
   {Our models explain the
     observed soft spectra and peak X-ray luminosities of individual superbubbles. 
     Each supernova event inside a superbubble
     	produces a fairly similar heating-entrainment-cooling
        sequence, and the energy content of superbubbles is
        always determined by a specific fraction of the energy released by one supernova.
	For a given superbubble, soft X-rays trace the internal energy content well with moderate scatter.
        Some mechanism seems to delay the energy loss in real
        superbubbles compared to our simulations. Alternatively, some mechanism
	other than \rv{thermal emission of} superbubbles may contribute 
	to the soft X-ray luminosity of star-forming galaxies. 
     }

   \keywords{Galaxies: ISM -- ISM: bubbles -- ISM: structure --
  hydrodynamics -- Instabilities -- X-rays: ISM
               }

   \maketitle
%

\defcitealias{Krausea13a}{Paper~I}

\section{Introduction}\label{sec:intro}
Stellar feedback is an essential ingredient in galaxy evolution models
\citep[e.g.][]{SGP03,Scanea08,EB10,PS11,Scanea12,Henrea13,Romea13}:
it regulates the star-formation rate, 
mass concentration and angular momentum distribution, the scale height of 
the interstellar medium 
\citep[e.g.][]{dAB05,DBP11b}, and the chemical evolution of stars and 
gas \citep{BW10,Putmea12}. Predictions are, however, not 
based on feedback calculations from first principles, and different prescriptions for 
stellar feedback lead to major deviations in the properties of simulated galaxies \citep{EB10,Scanea12,Agea13}: 
 the resulting total stellar mass is uncertain by a factor of a
 few; stars tend to be too concentrated for all prescriptions, which
 leads to unrealistically declining rotation curves; even the
 morphological type of a galaxy, i.e. whether or not it has a stellar
 disc, seems to depend on the feedback implementation.  It is clear
 that stellar feedback is important for the evolution of all but perhaps the
 most massive galaxies, where supermassive black holes may
 dominate \citep[e.g.][]{Krause2005b,Crotea2006,Nesea08,KG10,Gaspea12,SM12}.

Massive stars are the main agents of stellar feedback and form mainly in groups
\citep{ZY07}. 
Their energy output produces bubbles, shells and bipolar structures, which are observed 
in great detail in nearby star-forming complexes \citep[e.g.][]{Nielsea09,Mottea10,Preibea12,Miniea13,Russea13}. 
The compression of surrounding gas may 
trigger further star formation \citep{Ohlea12,Roccea13}.
Star-forming regions such as the Carina, Cygnus or Orion star-forming complexes form
a total of around 100 massive stars ($>8 M_\odot$) 
each \citep{Knoea02,Vossea10,Vossea12}. Population synthesis \citep{Vossea09} 
gauged by observations of the stellar content predicts
 the output of energy, which may be compared to the energy 
needed to create the observed bubbles. For example, in the case of Orion, the massive
stars have output about $2\times 10^{52}$~erg of kinetic energy into the surrounding gas, 
which agrees with the amount of energy needed to create the Orion-Eridanus superbubble \citep{Vossea10}. Such superbubbles are prominent structures in the interstellar
medium of the Milky Way and nearby galaxies with features 
e.g. in molecular gas \citep{DawsonJea13},
HI \citep[so-called HI-holes,][]{Boomea08,Bagea11,EP13},
H$\alpha$ \citep[][]{Rossaea04,Voigtea13}, 
Gamma-ray lines \citep[radioactive trace elements ejected from massive stars,][]{Knoea02,Diehlea06,Diehl13,Kretschea13},
and Gamma-ray continuum due to cosmic rays \citep{Ackea11}.

Uncertainties are substantial in the study of feedback effects: for example, recent 
{\it Herschel} observations suggest that the number of massive stars may have been underestimated by up to a factor of two 
in the Carina nebula complex \citep{Roccea13}, uncertainties 
from stellar evolution calculations and wind prescriptions lead to an uncertainty
of the energy output of tens of per cent \citep{Vossea09,Vossea12}. Uncertainties also exist in the coupling
of the energy output of the massive stars to the ambient ISM: 
hydrodynamic simulations show that 
the major fraction of the injected energy is lost to radiation 
\citep[e.g.][]{Tenea90,Thornea98,FHY06,CTB13} 
and that the clustering may affect the energy retained in the gas by a factor of a few \citep[][hereafter Paper~I]{Krausea13a}.

In order to improve the accuracy of feedback modelling,
it is therefore important to exploit all the available  observational constraints.
One such constraint is the diffuse soft X-ray emission, on which we focus here.
\rv{In general, diffuse X-ray emission from star-forming regions may be associated
with unresolved stars, cosmic ray acceleration and hot gas. Because superbubbles
are much larger than the associated energy-liberating star-forming region, unresolved
stars are only important where the star-forming complex is observed directly 
\citep[e.g.][]{Munoea06}. Cosmic rays are evidently accelerated in superbubbles.
This follows directly from the \fermi detection of diffuse cosmic ray emission in the 
Cygnus superbubble \citep{Ackea11}, and is suggested by the high-energy detections of 
star-forming galaxies \citep[e.g.][]{Ackea12} and the infrared/radio correlation
\citep[e.g.][]{GA09,SB13}. Cosmic rays are expected to take a share of a few, up to ten 
per cent of the energy injected by massive stars via diffusive shock acceleration
\citep[][for recent reviews]{Schurea12,Vink12,Bell13,Riegea13}. This process may work for 
shocks from winds and supernovae in various environments \citep[e.g.][]{Ellisea12},
including superbubbles \citep{Parizea04,BB08,BGO13}.  
Non-thermal X-ray spectra due to cosmic rays are detected
in the diffuse emission associated with some massive star clusters, e.g. 
\object{30~Doradus} \citep{Bambea04} and \object{Westerlund~1} \citep{Munoea06}.
However, many superbubbles do not show evidence for non-thermal X-ray emission
\citep[e.g.][]{Yamea10}. Where observed, the non-thermal emission is more important
above 2~keV, and softer, thermal components may often be extracted from the 
X-ray spectra \citep[e.g.][]{Bambea04,Munoea06}. Hot gas
emits diffuse soft X-ray emission in individual superbubbles \citep[e.g.][]{Sasea11,KSP12}
and entire star-forming galaxies \citep{Strickea04a}. Here, we restrict our attention 
to the diffuse, soft X-ray emission due to hot gas, noting that Cosmic rays are not explicitly 
accounted for in our analysis.}
%
%
\begin{table}
  \caption{Simulation parameters}             
  \label{t:simpars}      
  \centering          
  \begin{tabular}{l r r r  r}     
    \hline\hline       
    Label$^a$ & Star mass /$M_\odot$ & X$^b$ / pc & Y / pc & Z / pc \\
    \hline                    
    S25     &  25  &  0  &  0  &  0  \\
    S32     &  32  &  0  &  0  &  0  \\
    S60     &  60  &  0  &  0  &  0  \\
    3S0     &  25  &  0  &  0  &  0  \\
    &  32  &  0  &  0  &  0  \\
    &  60  &  0  &  0  &  0  \\
    3S1     &  25  &  -30  &  10  & 10  \\
    &  32  &  -25  &  -10  &  0  \\
    &  60  &  0  &  0  &  0  \\
    3S2     &  25  &  -60  &  20  &  10  \\
    &  32  &  50  &  -10  &  0  \\
    &  60  &  0  &  0  &  0  \\
    \hline                  
  \end{tabular}
\tablefoot{$^a$Label of respective run. Without addition, the run labels refer to the 2~pc resolution runs. Run 3S1 has also been carried out at 1~pc resolution, which is denoted by the extension '-mr'. Runs 3S0 and 3S1 have also been carried out  at 0.5~pc resolution, denoted by the extension '-hr'. $^b$ X, Y, and Z denote the position of the stars.} 
\end{table}
%


Even wind bubbles of isolated massive stars should be expected to 
\rv{produce hot gas} 
due to the high wind velocities \citep[$>1000$~km~s$^{-1}$, e.g.][]{Pulsea96,VG12},
especially in the high-power Wolf-Rayet phase \citep{Graefea11},
and thus to be X-ray bright.
\rv{Indeed} two such examples are known \citep{ZK11,Toalea12}.
The interaction of individual bubbles leads to
the formation of superbubbles 
\citep[for reviews]{OCM01,Chu08,Oey09}: 
they reach sizes of several 100~pc. When they are X-ray-bright this is suspected to be related to recent supernova activity. 
Superbubbles \rv{usually} have a thermal X-ray spectrum
\citep[e.g.][\rv{compare above for non-thermal contributions}]{Sasea11,Jaskea11} with typical temperatures around 0.1 keV and luminosities
of the order of $10^{35}-10^{36}$~erg~s$^{-1}$. 
They are often surprisingly bright compared to expectations from models \citep{OG04},
which is usually explained by entrainment of mass due to interaction with the shell walls \citep{Jaskea11}.
Hydrodynamic instabilities play a key role in simulations of superbubbles
\citep[e.g.][]{BdA06}, but have so far not been quantitatively assessed regarding 
their effect on the general X-ray properties of superbubbles.


In a previous paper \rv{(Paper I)}, we simulated superbubbles emerging from 
three massive stars in a constant 
density environment. We plan to compare the results in detail to 
superbubble observations in various wavelength regimes. Here, we focus on
X-rays: a key observational feature, which we reproduce well, is the 
large X-ray variability. A recent supernova may boost the X-ray luminosity by a factor of up to a hundred for a timescale of order $10^5$~yr. Regarding morphology, we compare to previously unpublished data from the nearby Orion-Eridanus superbubble. We interpret the observed systematic spectral variations with position on the sky as global oscillations of the hot gas inside the bubble. The temperature-luminosity
diagram is compared to data from superbubbles in the Large Magellanic Cloud (LMC). 
While the general ranges of luminosities and temperatures show reasonable overlap,
there are also some discrepancies.
We then compare the total \rv{thermal} X-ray emission integrated over the lifetime of a simulated 
superbubble as a fraction of the input energy to the corresponding number derived 
from the soft X-ray luminosity and the star-formation rate for nearby galaxies.
The simulations underpredict the data by a factor of ten. We discuss possible reasons for this in Sect.~\ref{sect:disc}.

%
\section{Simulations and their analysis}\label{sect:am}
Our analysis is based on 3D adaptive-mesh-refining hydrodynamics 
simulations with the {\sc Nirvana}-code 
described in detail in Paper~I. In short,
we follow the evolution of the 100~pc scale circum-stellar medium of a group of three massive stars (25, 32 and 60 $M_\odot$) for the entire evolution of these stars, including the final supernova-explosion
(at 8.6, 7.0 and 4.6 Myr, respectively), and a few Myr beyond. 
\rv{We take into account radiative cooling and heating, but no non-thermal processes.}
The individual bubbles merge and form a superbubble at around 1~Myr of simulation time. This superbubble features a cool shell, which dissipates most of the energy radiatively, and a hot bubble interior at sufficiently high temperature to emit thermal X-ray radiation. We varied the spatial configuration of the three massive stars (compare Table~\ref{t:simpars})
from cospatial (3S0) over tens of parsecs apart (3S1, 3S2), 
to very large distances, realised by having each star in a separate simulation 
(S25, S32 and S60) and adding the result. 
We performed all runs at a finest resolution of 2~pc, and repeated two (3S0, 3S1) at higher resolution, up to 0.5~pc.
The resolution comparison shows that the X-ray properties are well converged in the X-ray-bright phases. Details are discussed
in Appendix~\ref{a:rescomp}.

We followed \citet{BdA06} in assuming that resolved hydrodynamic instabilities 
are the main driver of mixing within the bubbles, and consequently neglected 
any evaporation due to thermal conduction. Our analysis therefore excludes 
the main-sequence phase of the most massive star, where the wind is too steady for 
prominent instabilities to develop.  X-ray spectra at the relevant 
temperatures (one to a few million Kelvin) are dominated by a large number of emission lines. 
We used a model for emission from tenuous hot plasma which includes the relevant atomic shell physics.
Such a model, called 'Mekal' \citep{MGvdO85,MLvdO86,LOG95}, is conveniently provided via the {\sc XSPEC}-package v12.8.0\footnote{http://xspec.gsfc.nasa.gov/},
\citep{Arnaud96,DA01}. We produced spectra for gas at solar abundances and with temperatures
between 0.1 and 10~keV in steps of 0.1~keV, and convolved the spectra with the temperature
distribution of the hydrodynamic simulations. 

%
%
\begin{table*}
  \caption{X-ray luminosities at different characteristic times and energy bands}             
  \label{t:xlum}      
  \centering          
  \begin{tabular}{l l l l l l l l l}     
    \hline\hline       
    Time / Myr & Description & \multicolumn{7}{c}{Log (luminosity / (erg s$^{-1}$))} \\
	&	& 0.1-20 keV & 0.2-12~keV & 0.2-0.5~keV & 0.5-1~keV&1-2~keV & 2-4.5~keV & 4.5-12~keV \\
   \hline   
	4.5251	& First WR & 34.47& 34.27 & 34.11 & 33.71 & 32.61 & 31.78 & 31.43 \\
	4.6016	& First SN  &  34.36 &34.17 & 33.99 & 33.59 & 32.67 & 32.46 & 32.55 \\
	4.6159 	& First X-ray max. & 36.12 &  35.93 & 35.77 & 35.38 & 34.32 & 33.30 & 32.72 \\      
	9.2236	& Fading superbubble & 33.71 & 33.53 & 33.34 & 33.02 & 32.18 & 31.34 & 30.56 \\         
    \hline                  
  \end{tabular}
\end{table*}
%

%
\section{Results}\label{sect:res}

\subsection{X-ray emission in different evolutionary phases}
\begin{figure*} 
   \resizebox{.97\hsize}{!}
            {\includegraphics{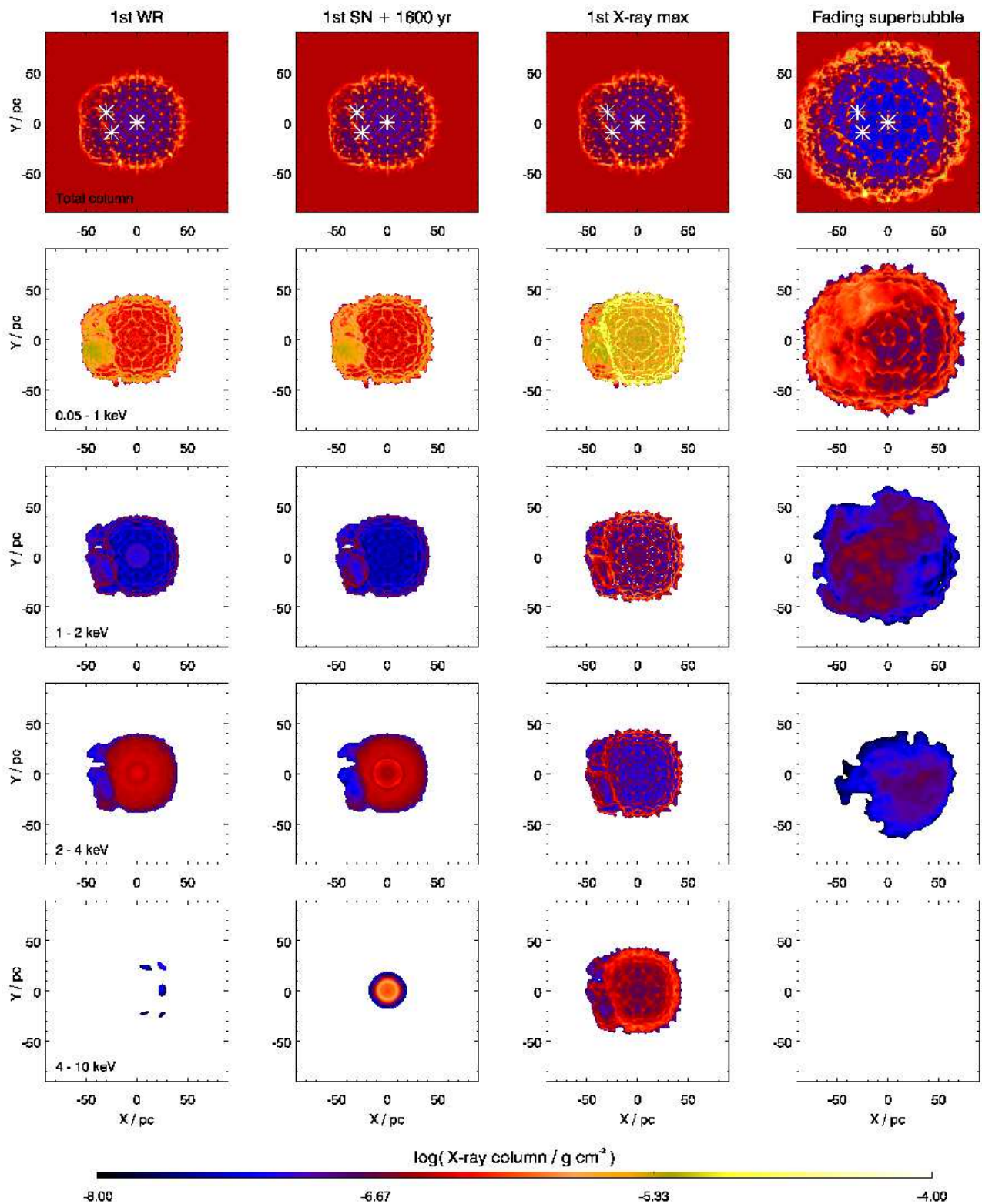}}
      \caption{Time evolution (from left to right; the snapshot time is indicated on the top of each column) 
		of X-ray properties of simulation 3S1-hr, featuring 3 massive stars
              	at about 30~pc distance from each other. From top to bottom, the plots 
		show column density distributions of all gas (the three massive stars are indicated as white stars), 
		and of X-ray-emitting gas at temperatures of 0.05-1~keV, 1-2~keV, 2-4~keV and 4-10~keV. The colour bar gives 
		values relevant to the X-ray panels, but the colour scale is the same for the total gas column.
		The left column shows the superbubble in the Wolf-Rayet phase of the central, most massive star.
		The middle columns display two different snapshots shortly after the first supernova, first with prominent
		high energy and then with very strong soft emission, when the shock approaches the shell. Right column:
		0.6~Myr after the final supernova, which occurred off-centre, the hot gas performs global oscillations, which leads 
		to the different morphologies in the different bands. See Table~\ref{t:xlum} for exact snapshot times and 
	corresponding X-ray luminosities. A movie is provided with the online version.}
         \label{fig:overview}
   \end{figure*}

\begin{figure}
   \includegraphics[width=\hsize]{\fdir/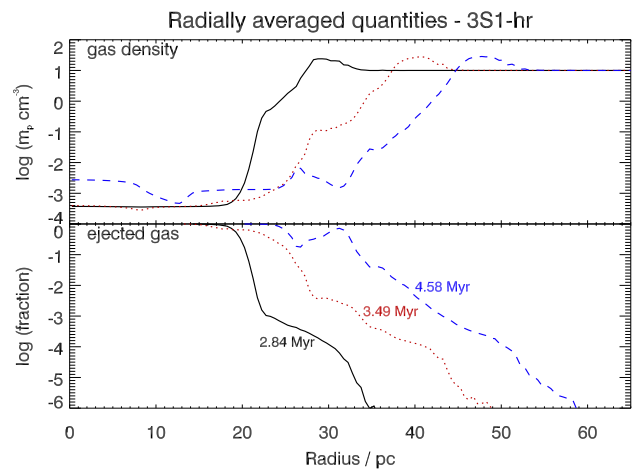}
       \caption{Density (top) and fraction of ejected gas as a function of distance from the 
       	most massive star (radius) for three different times (solid black: 2.84~Myr, dotted 
	red: 3.49~Myr, dashed blue: 4.58~Myr), averaged at constant radius.
	In the main-sequence phase of the 60~$M_\odot$ star (2.84Myr), there is a sharp
	contact surface at 20~pc, where the ejected gas fraction suddenly drops to $10^{-3}$.
	During the Wolf-Rayet phase, the wind power increases. Rayleigh-Taylor instabilities
	at the accelerating shell create an extended mixing layer. 20\,000~yr before the 
	supernova of the 60~$M_\odot$ star	 (dashed blue curves), the mixing layer extends 
	between about 27 and 40~pc, if defined by ejected gas fractions between 1 and 10
	per cent.
              }
         \label{fig:mixing}
   \end{figure}

The contribution
to the different parts of the X-ray spectrum depends on the exact 
temperature of the gas. 
We show the temporal evolution of the column density of gas in four different temperatures ranges for
the representative run 3S1-hr (Table~\ref{t:simpars} for details) in Fig.~\ref{fig:overview}. 
For each phase, the synthetic X-ray luminosity is provided in representative bands 
in Table~\ref{t:xlum}.
The maps differ markedly 
in the different evolutionary phases 
of the superbubbles: as discussed in Sect.~\ref{sect:am}, we do not analyse the phase when all three stars are still on the main sequence.
In the first Wolf-Rayet phase (Fig.~\ref{fig:overview}, first column), the shell is 
accelerated, and strong Rayleigh-Taylor instabilities lead to an extended 
mixing layer (Fig.~\ref{fig:mixing}) at temperatures suitable for soft X-ray emission.
The luminosity is highest at the low-energies ($<1$~keV, $>10^{34}$~erg~s$^{-1}$).
After merging of the individual wind shells, the debris of the shell interfaces 
is being pushed into the lower pressure regions, i.e. the bubbles of the less massive stars.
The eroded shell fragments mix with the gas that pushes them, 
which again leads to a substantial mass of sub-keV gas.
For the main bubble, the amount of sub-keV gas is
larger towards the edge of the bubble, as the mixing layer  is located outwards of about 25~pc in this phase (compare 
Fig.~\ref{fig:mixing}). The network produced
by the Vishniac instability features prominently, as mixed, warm gas is focused in the small 
cavities in the shell (compare Fig.~3 in Paper~I).
The 1-2~keV emission is produced almost exclusively in the mixing zone, inside of the
sub-keV emission. The unmixed Wolf-Rayet wind itself fills the inner parts of the bubble
with hot gas at temperatures between 2~and 4~keV. The total emission is however $<10^{32}$~erg~s$^{-1}$
in the 2-4.5~keV band, and therefore hardly expected to be detected in observations.
We find almost no gas above 4~keV. Note, that we do not treat cosmic ray acceleration, which 
would be expected in such conditions, and which may produce non-thermal emission at a few keV 
\citep[compare e.g. discussion in][]{Yamea10}.
%
%
%

The second column in Fig.~\ref{fig:overview} shows the situation  
1600~yr after the first supernova explosion. The 
supernova has produced a substantial amount of high temperature gas
which has boosted the emission in the high-energy bands by about an order of magnitude
(compare second row in Table~\ref{t:xlum}). 
There is no change in the distribution of the colder gas,
because the shock wave has not yet reached the higher density regions, and consequently,
there is no change to the low-energy bands.

About 14\,000~yr later (third column in Fig.~\ref{fig:overview}), the shock wave has 
traversed the entire mixing layer. Because the density increases outwards 
(Fig.~\ref{fig:mixing}) the shock slows down, with less heating 
in the outer regions.
Consequently, the emission spectrum becomes softer. The overall emission, in particular the 
sub-keV band, has increased by more than a factor of 40 compared to times before the
supernova (third row in Table~\ref{t:xlum}).

The X-ray emission fades on a timescale of several $10^5$-$10^6$~yr 
after each supernova, faster in the higher energy bands. This is due to 
enhanced mixing: when the shock wave 
reaches the shell, strong Rayleigh-Taylor instabilities are triggered, 
mixing entrained gas with the hot gas, thereby reducing its temperature. The acceleration of 
the shell also leads to some adiabatic-expansion losses, which reduce the X-ray luminosity.
We show the superbubble 0.6~Myr after the third explosion in the right column of Fig.~\ref{fig:overview},
where the X-ray emission has faded below earlier phases (fourth row in Table~\ref{t:xlum}).
This supernova exploded off-centre and produced a global oscillation
of the gas in the bubble interior. At the time shown, the left part is expanded and hence colder, whereas the right part is compressed and therefore hotter. There is no gas at temperatures above 4~keV at this time.

\subsection{Spectral evolution}
\begin{figure*} 
   \resizebox{\hsize}{!}
            {\includegraphics{\fdir/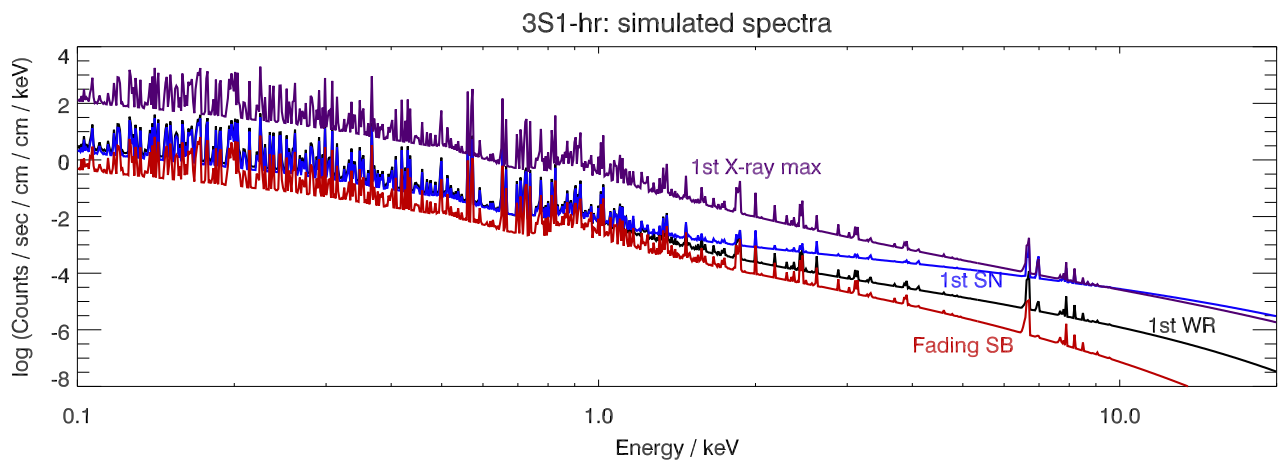}}
      \caption{Thermal X-ray spectra for run 3S1-hr. The time sequence 
	of the labels is \rv{1st~WR (black), 1st~SN (blue), 1st~X-ray-max (violet), and
	Fading SB} (red). They 
	refer to the descriptions in Table~\ref{t:xlum}. A supernova inside a superbubble
	first increases the hard 
	X-rays and then the softer parts by up to two orders of magnitude compared to 
	immediately before the explosion. All the spectra reflect the multi-temperature structure.}
         \label{fig:spectra}
   \end{figure*}

Spectra for the entire superbubble of run 3S1-hr are shown in Fig.~\ref{fig:spectra}.
At no time, we find a strong cutoff, as would be the case if one temperature would 
strongly dominate the emission. Instead, the contributions of regions with different 
temperatures add up to a shape that comes close to a broken power law. Up to 0.5~keV,
fitting gives a power law index close to -1.5. For the energy range 0.5-20 keV, the 
fitted power law indices are between -4.4 and -2.7. Discrepancies from single temperature
spectra in collisional ionisation equilibrium are also found in observations
\citep{Jaskea11,Sasea11,KSP12}. At the given temperatures, departures from
collisional ionisation equilibrium are small \citep{SD93}. Hence, a multi-temperature 
interpretation of the observations is very probable in agreement with our findings.

\subsection{Spatial configuration of the massive stars}
\begin{figure*}
   \centering
   \includegraphics[width=\hsize]{\fdir/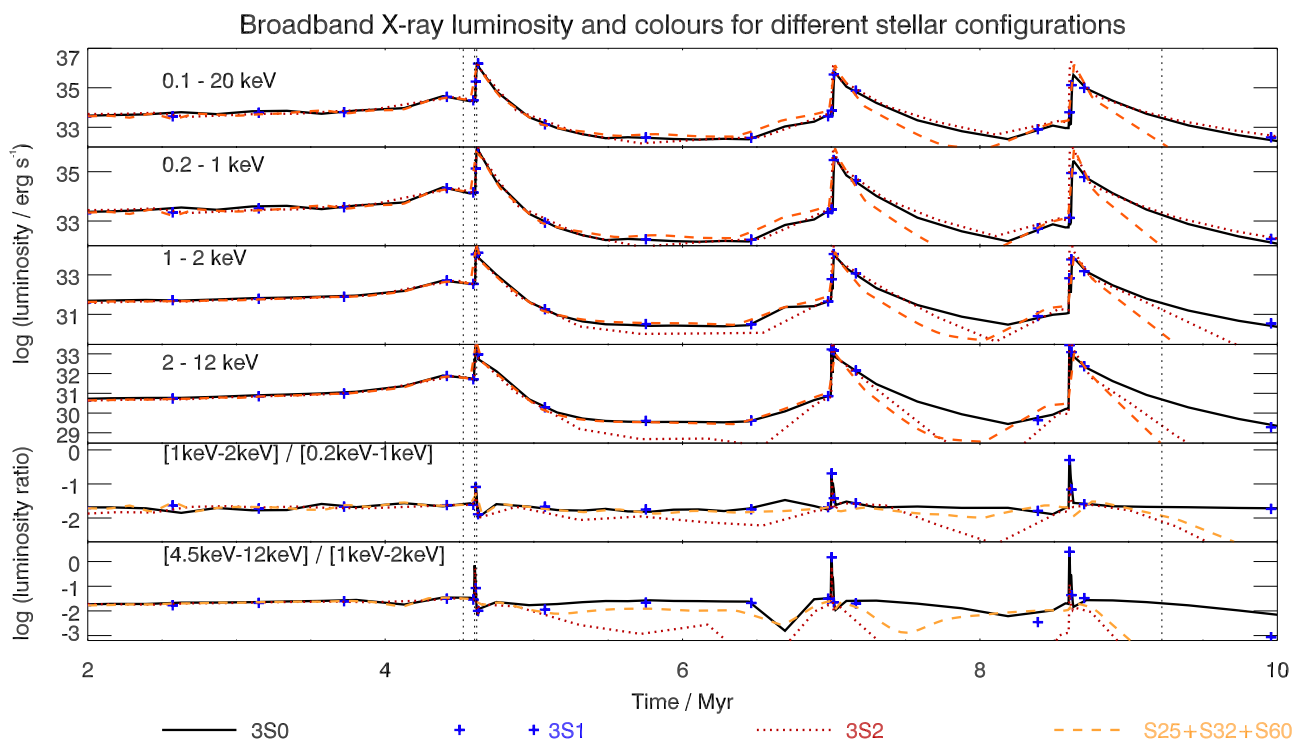}
       \caption{Integrated thermal X-ray luminosity for four energy bands
	(four top panels, energy bands indicated in the respective panel) and two 
	X-ray colour indices (bottom panels) as a function of time since the coeval formation of 
	the three massive for different spatial configurations of these stars. The configurations 
	of the 3 massive stars are: all stars at the same place (3S0, solid black), 
	as indicated in Fig.~\ref{fig:overview} (3S1, blue pluses), at a significantly 
	larger distance (3S2, red dotted) and all stars in separate simulations
	 (dashed orange line). The vertical dotted lines indicate the snapshot times 
	used for Fig.~\ref{fig:overview}.
	All runs show a strong variability in all X-ray bands. Clustering increases 
	the X-ray output long after a supernova event (peaks) and hardens the 
	spectrum. Both colour indices stay close to -1.7 for most 
	of the time. See text for more details.
              }
         \label{fig:conf}
   \end{figure*}

We investigated different configurations of the stars, varying the distance 
from co-spatial
over 2 configurations with tens of parsecs distance to very large distances 
(separate simulations for each star). 
For this comparison, we took only simulations with a resolution of 2~pc.
The X-ray luminosity, especially the hard bands, fades faster, as the stellar 
distances increase (Fig.~\ref{fig:conf}). This is, because we approach the isolated bubbles case:
a supernova exploding in a smaller bubble 
produces a higher pressure, hence a stronger shell acceleration and thus 
more mixing via the Rayleigh-Taylor instability and faster adiabatic losses. 

We defined two colour indices as the ratio of the emitted power in bands as defined 
in Table~\ref{t:xlum}:
\begin{eqnarray*}
C1&=&\log_{10}\nbr{\sbr{1-2\,\mathrm{keV}}/\sbr{0.2-1\,\mathrm{keV}}}\\
C2&=&\log_{10}\nbr{\sbr{4.5-12\,\mathrm{keV}}/\sbr{1-2\,\mathrm{keV}}}
\end{eqnarray*} 
For the strong clustering runs, both, $C1$ and $C2$, 
have values close to -1.7 throughout the evolution, except shortly after each supernova. 
As we placed the stars further away from each other,
both, $C1$ and $C2$ dropped substantially.

The X-ray luminosity directly associated with the supernova explosions, 
until a few $10^5$~yr after the explosion, varies remarkably little in the different bands.
Even the final supernova produces the same peak luminosity in all bands, 
irrespective of the supernova exploding in its own wind bubble or the cavity 
being shaped by the complete 
evolution of two more massive stars at different positions. We find a peak X-ray luminosity 
of roughly $10^{36}$~erg~s$^{-1}$ after each supernova for all our configurations.
%

The increase in the colour indices after each supernova explosion is significant and the 
decay back to the base level is well resolved in time: $C1$ increases to up to~-0.5 and $C2$ to
positive values. 

In summary, unless a supernova has just exploded, clustering leads to higher X-ray luminosities
in all bands and harder colour indices. When a supernova actually explodes, the X-ray luminosity
is always increased to the same level in all bands.

%

   \begin{figure}
   \centering
      \includegraphics[width=\hsize]{\fdir/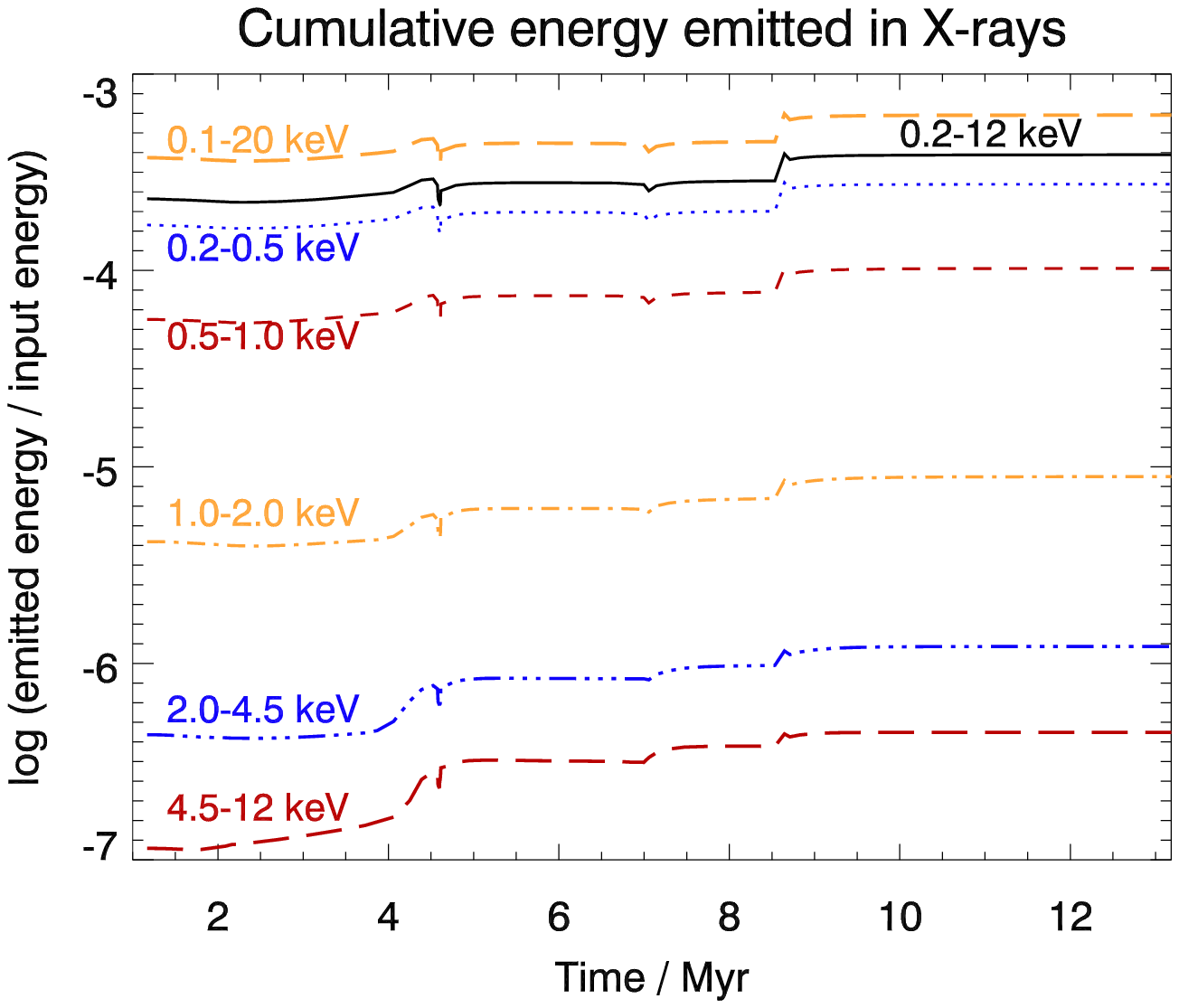}
      \caption{Cumulative energy output as fraction of the cumulative 
	\rv{energy input over time}
      	for various energy bands indicated next to the respective curves and for run 
	3S1-hr, shown in Fig.~\ref{fig:overview}. 	
              }
         \label{fig:ecum}
   \end{figure}
%

\subsection{Total energy emitted in X-rays}
The time-integrated radiated energy 
as a fraction of the current cumulative input energy 
is shown in Fig.~\ref{fig:ecum}. For each energy band, the emission 
stays at the same value within an order of magnitude, throughout the evolution. 
The ratio between energy bands is almost constant. The curves 
decrease after the first supernova, but quickly return to a similar level, 
as the injected energy is emitted when the shock reaches the shell. 
The curves then \rv{stay} constant until shortly before 
the next supernova. Generally, the emission drops
by a factor of a few between the 0.2-0.5 and the 0.5-1~keV bands, a further factor of 
ten for the 1-2~keV band, and another factor of ten to the higher energy bands.
The total emitted energy in X-rays is a  few times $10^{-4}$ of the injected energy.
This result can be used to predict the total X-ray luminosity of nearby galaxies 
from such superbubbles (sect.~\ref{ssect:nearbygals}, below).

   \begin{figure}
   \centering
  	    \includegraphics[width=\hsize]{\fdir/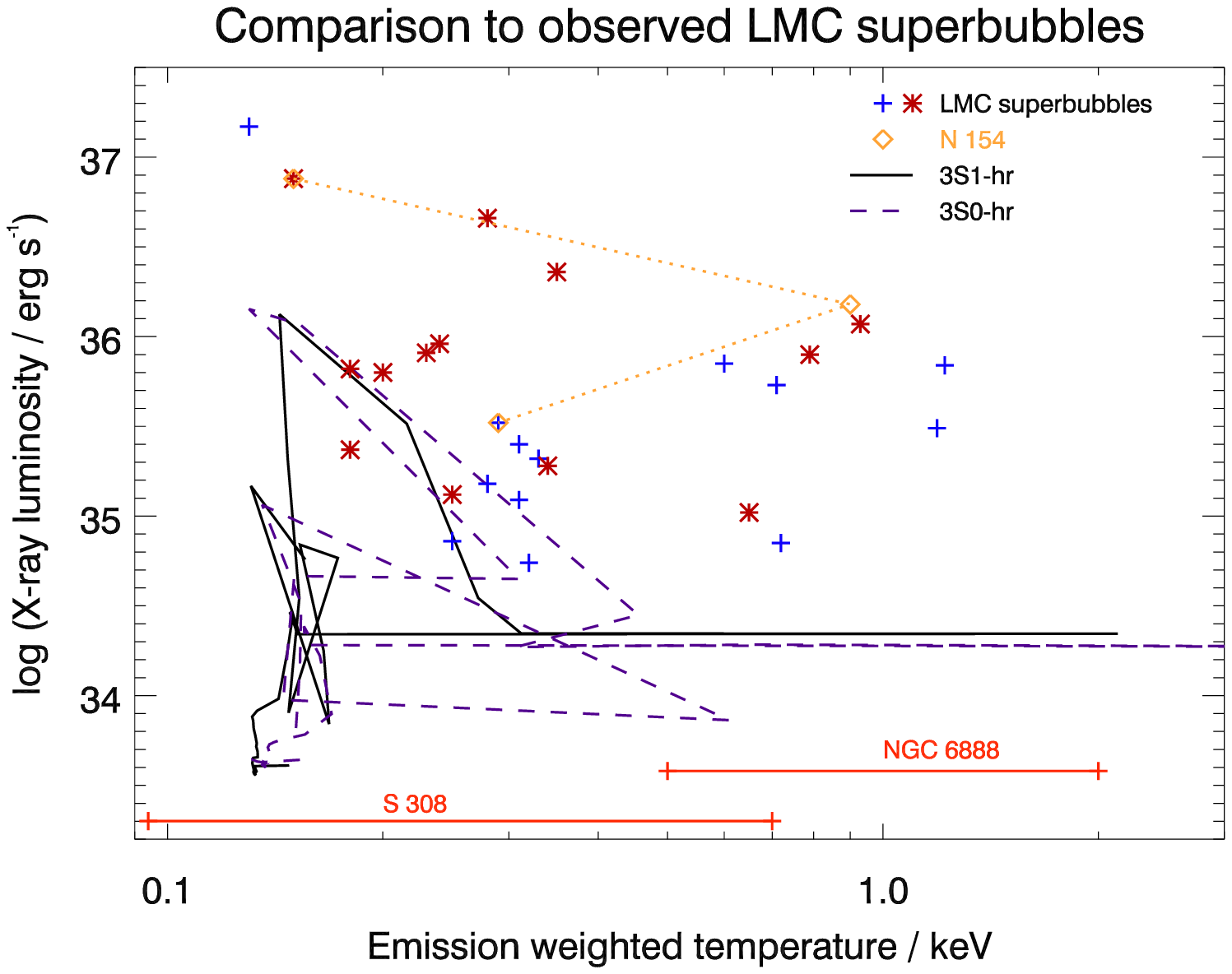}
      \caption{Luminosity-temperature diagram for run 
	3S1-hr (solid black line), shown in Fig.~\ref{fig:overview},
	and for run 3S0-hr, where the massive stars are all co-spatial.
	For this comparison, the X-ray luminosity is integrated between 0.1 and 2.4~keV
	corresponding to the \rosat~PSPC instrument. The temperature is the average temperature
	of the superbubble weighted by the emission measure.
	This is compared to the data of
	a sample of LMC superbubbles observed with
	{\it \rosat} by \citet[][red stars and the blue pluses correspond
	to the same sample but different absorption corrections]{DPC01}.	The light-orange
	diamonds connected by the dotted line correspond to different measurements
	for the bubble N~154 in the LMC. The two lower temperature measurements are from \rosat. 
	The higher temperature measurement is from \citet{Sasea11} with {\it \xmm}.
	This gives an idea of the observational uncertainty.
	We also show data for the two X-ray-detected Wolf-Rayet bubbles, S308 \citep{Toalea12}
	and NGC~6888 \citep{ZK11}.
              }
         \label{fig:emwt}
   \end{figure}
%

\section{Comparison to observations}\label{sect:obs}

\subsection{General X-ray properties of superbubbles}\label{ssect:obs-gen}
\citet{DPC01} present a {\it \rosat} study of 13 superbubbles in the 
LMC. The superbubbles have diameters of about 100~pc, very similar to our 
simulations. They find diffuse X-ray emitting regions in each case,
with a soft spectrum  
and a patchy morphology, sometimes outside apparent shells as defined by H$\alpha$.
They report correlations of the X-ray luminosity with the H$\alpha$ luminosity,
the expansion velocity of the shells, and the OB star count. Apart from the correlation 
with the OB star count (which we do not address here, 
because all our simulations have the same three stars) all these findings 
are well explained by our simulations: Whenever two bubbles merge, the pressure
in each individual bubble is likely very different. In our simulations (Fig.~\ref{fig:overview}),
the emission is enhanced in the gas flooding the low-pressure bubble, because
the bubble interface is eroded. Such gas should therefore be highlighted in observations.
We find the gas to become X-ray bright, when entrained and mixing shell gas is shocked 
due to a supernova. This is necessarily connected \rv{to} an acceleration of the shell,
in line with the observed correlation between X-ray luminosity and expansion velocity.
The enhanced expansion velocity then boosts the H$\alpha$ luminosity at the leading shock.

\citet{ZK11} present a {\it SUZAKU} study of the Wolf-Rayet wind bubble \object{NGC~6888}.
They find emission from a variety of temperatures with a dominant 
low-temperature component, but also 
contributions from gas above 1~keV. The emission is limb-enhanced and originates
in clumps. This agrees very well with our simulations, which show prominent Rayleigh-Taylor
instabilities in the Wolf-Rayet phase entraining filaments from the shell into the bubble.
The X-rays would then be produced in the mixing region.

We compare luminosities and emission-measure weighted temperatures from our 
high-resolution simulations 
to these observational data in Fig.~\ref{fig:emwt} (the tracks of the other ones are similar). 
Here, we set the X-ray luminosity to zero below $10^{33.5}$~erg~s$^{-1}$, because for lower 
luminosities we expect either to be dominated by heat conduction effects (main sequence phase),
which are not included in the simulations, or the simulations are not numerically converged
(compare Fig.~\ref{fig:3S1rescomp}). The plot shows that the values we derive are 
generally in the range of expectations. Our simulation results cover the 
full range of observed bubble temperatures. Depending on the assumptions about absorption
in the observations, they also cover part or most of the range of observed luminosities.
In detail, the LMC superbubbles seem to require somewhat hotter and at the same time
more luminous gas than what we find in our simulations. 
Also, the duration of the X-ray luminous phase of order $10^5$~yr (Fig.:~\ref{fig:conf})
appears short. It would mean that only few superbubbles are
X-ray luminous. While we are not aware of a study that analyses the X-ray luminosities
of an e.g. optically selected parent sample, this number appears to be small.

For the Wolf-Rayet winds, we find lower luminosities ($\approx 10^{34}$~erg~s$^{-1}$) than for superbubbles. This agrees well with observations (also Fig.~\ref{fig:emwt}).

\subsection{The Orion-Eridanus bubble}\label{ssect:obs-oeb}
\begin{figure}
   \centering
     \rotatebox{-90}{\includegraphics[width=\hsize]{\oebdir/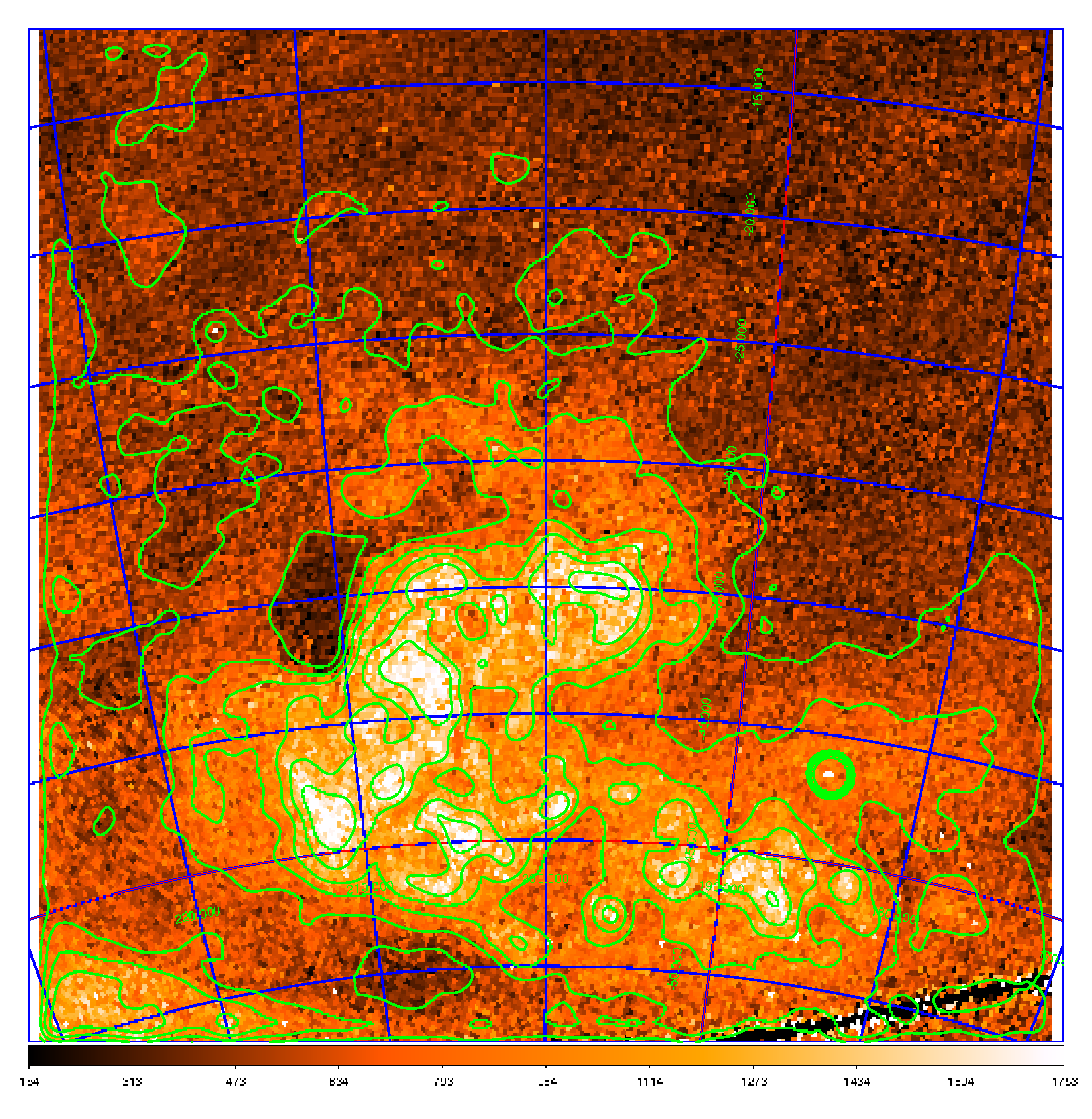}}
     \rotatebox{-90}{\includegraphics[width=\hsize]{\oebdir/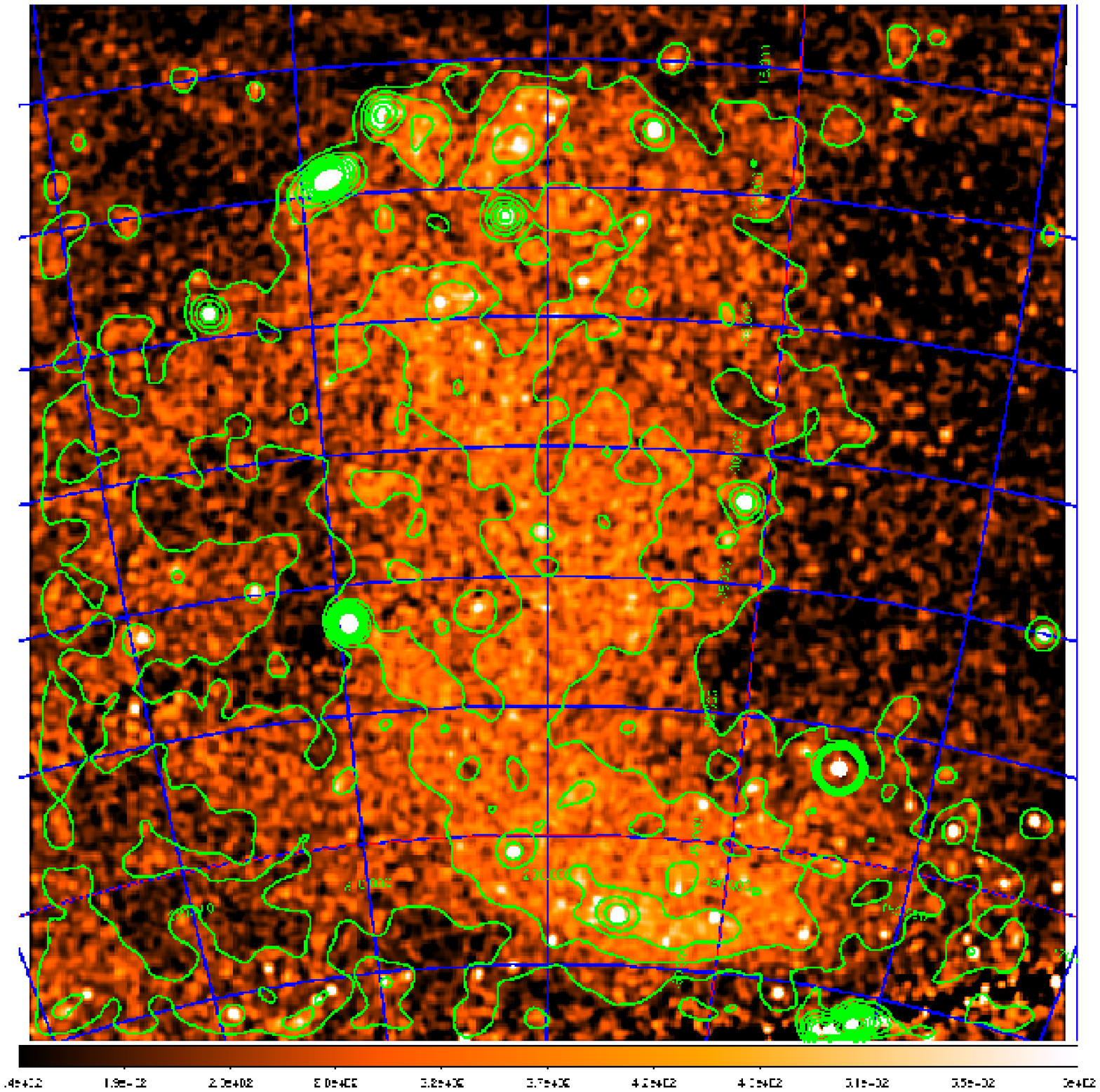}}
      \caption{X-ray maps of the Orion-Eridanus superbubble 
	(same region as in Fig.~\ref{fig:OEB-Ha}) in the 
	0.1-0.4~keV (top, contour levels: 700,900,1100,1300,1500,1900) and the 0.5-2~keV 
	(bottom, contour levels: 220,300,380,460,540,620,700) bands from the \rosat
	all-sky survey \citep{Snowdea97}. The colour scale 
	units are $10^{-6}$~counts~arcmin$^{-2}$~s$^{-1}$. The scale is linear and 
	does not start at zero in order to de-emphasise unrelated background.  
	The X-ray emission of the superbubble interior is delineated by shells seen in H$\alpha$
	(compare Fig.~\ref{fig:OEB-Ha}). The softer and harder X-ray bands emphasise gas 
	at different temperature. We propose that this may relate to a global oscillation of the
	superbubble, similar as seen in the simulations.
              }
         \label{fig:OEB-X}
   \end{figure}

\begin{figure}
   \centering
     \rotatebox{-90}{\includegraphics[width=\hsize]{\oebdir/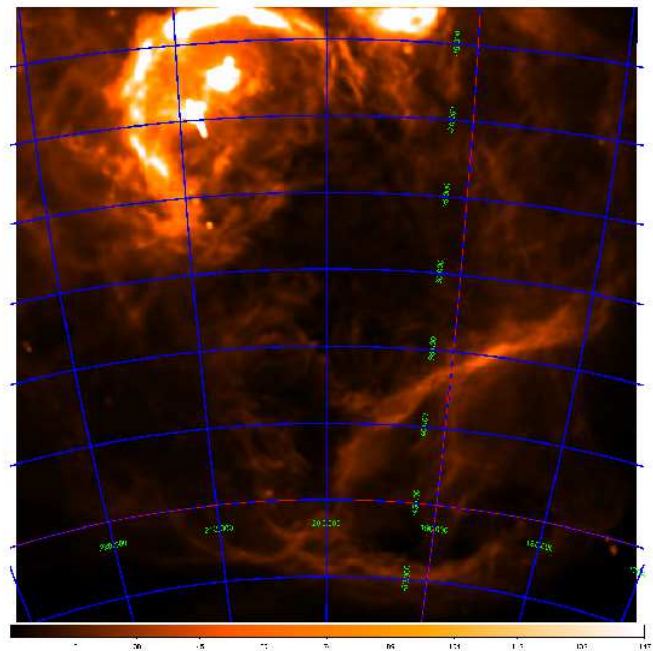}}
      \caption{H$\alpha$ map of the Orion-Eridanus superbubble \citep[][same region as in Fig.~\ref{fig:OEB-X}]{RO79}.
	The colour scale is linear and in units of Rayleigh 
	($4\pi\times 10^{-4}$~photons~cm$^{-2}$~s$^{-1}$~sr$^{-1}$).
       A galactic coordinate system is indicated with degree labels in green. 
	The projection is area-conserving 
	and centred on the Galactic anti-centre.	
	The two bright regions near $(l,b)=(-208,-18)$ are the Orion clouds.
	The bright arc towards the top left of the Orion clouds is Barnard's
	loop, probably a shell due energy injection from the Orion clouds. It delineates
	the extent of the superbubble towards the Galactic plane. The arc-like features
	towards the bottom of the map may also be related to the shell and delimit the 
        X-ray emission (compare Fig.~\ref{fig:OEB-X}).  The shell is also traced in HI
	\citep{BHB95}.}
         \label{fig:OEB-Ha}
   \end{figure}

The Orion-Eridanus cavity \citep[e.g.][Figs.~\ref{fig:OEB-X} and~\ref{fig:OEB-Ha}]{BHB95,Welshea05,Vossea10,Joea12} is a very
nearby superbubble, only a few hundred parsecs away. It extends about 40~degree on the
sky. It was discovered as a prominent source in the \rosat all-sky survey \citep{Snowdea97,FE99}.
\rosat continues to be the best source of soft X-ray data for such an extended
object. An analysis of the available \xmm data \citep{Lubos12} showed that more observing time
and dedicated pointings avoiding sources would be required 
to measure the extended X-ray emission.
The data have not yet been evaluated separately for the Orion-Eridanus
cavity. We show the \rosat images in two energy bands in Fig.~\ref{fig:OEB-X}. An H$\alpha$ map of the same region 
is shown in Fig.~\ref{fig:OEB-Ha}. The $1/4$-keV emission is bounded
by H$\alpha$ emission away from the Galactic plane (downwards in 
Figs.~\ref{fig:OEB-X} and~\ref{fig:OEB-Ha}). It fades towards the galactic plane as the
higher energy emission becomes more prominent. This is very similar to our simulated
morphologies at late times (compare Fig.~\ref{fig:overview}). 
In the simulations, the effect is due to global
oscillations of the hot gas in the superbubble. The Orion-Eridanus cavity has had
enough time for such global oscillations to establish: from the stellar population, one
expects about one supernova every Myr \citep{Vossea10}. 
The sound crossing time is about 0.5~Myr 
(150~pc, $c_\mathrm{s}=300$~km/s). This would suggest a 
significant probability to capture the superbubble in the second or 
third period of a global oscillation.
While a precise derivation of the luminosity
of the Orion-Eridanus superbubble has turned out to be difficult due to its proximity
and spatially varying absorption, it is likely that it significantly exceeds the 
$10^{34}$~erg~s$^{-1}$ (estimated from the counts in Fig~\ref{fig:OEB-X})
we typically measure for this phase. 
%

%

\subsection{Diffuse X-ray luminosity of galaxies}\label{ssect:nearbygals}
The soft ($<2$~keV) as well as the hard ($>2$~keV) X-ray emission correlates 
with tracers of the star-formation rate 
\citep{Persea04,Strickea04b}\footnote{\citet{Francea03} do not find 
a correlation between the star-formation rate 
and the thermal X-ray emission. However, from our spectra, we caution that due to 
the multi-temperature nature of superbubbles, it may be difficult to properly distinguish
thermal and non-thermal components from the observed spectra. Also, they only 
cover one dex in X-ray luminosity.}.
Our spectra (Fig~\ref{fig:spectra}) and the cumulative radiated energy plots (Fig~\ref{fig:ecum})
make clear that superbubbles do not contribute much -- thermally -- 
to the hard band, and indeed
the observed correlation in the hard band is thought to be due to the emission of 
high-mass X-ray binaries \citep[neutron stars with massive OB companions,][]{Persea04}.
Here, we assess the contribution of superbubbles to the soft X-ray emission of galaxies.

As shown above, a superbubble experiences strong changes in its
diffuse X-ray luminosity over time. For an entire galaxy, we can assume that we see
a large number of superbubbles in uncorrelated evolutionary states.
The ensemble average over all superbubbles in the galaxy 
would then be equal to the time average of the emission of one superbubble scaled 
by the galaxy's star-formation rate, \mbox{\it SFR}. Above (Sect.~\ref{sect:res} and Fig.~\ref{fig:ecum}),
we found for the conversion efficiency $f_\mathrm{sX}$
from mechanical power to diffuse soft X-ray luminosity 
a factor of a few times $10^{-4}$.  One supernova occurs approximately for every 
$100M_\odot$ of stars formed \citep[e.g.][]{Dahlea12}. There is a roughly 
equal energetic contribution from winds and supernovae, respectively, of $10^{51}$ erg 
averaged over the population \citep{Vossea09}, for each massive star.
These assumptions lead to a prediction for the average diffuse soft X-ray output of
of star-forming galaxies of 
\begin{equation}
L_\mathrm{sX} \approx 
\frac{2\times 10^{51}\mathrm{erg}}{100\,M_\odot}\, f_\mathrm{sX} \,\mbox{\it SFR}
=10^{38} \,\mathrm{erg\,s^{-1}} 
\frac{\mbox{\it SFR}}{M_\odot \mathrm{yr}^{-1}}
\end{equation}
from our simulations (0.2-2~keV band).
   \begin{figure}
   \centering
     \includegraphics[width=\hsize]{\fdir/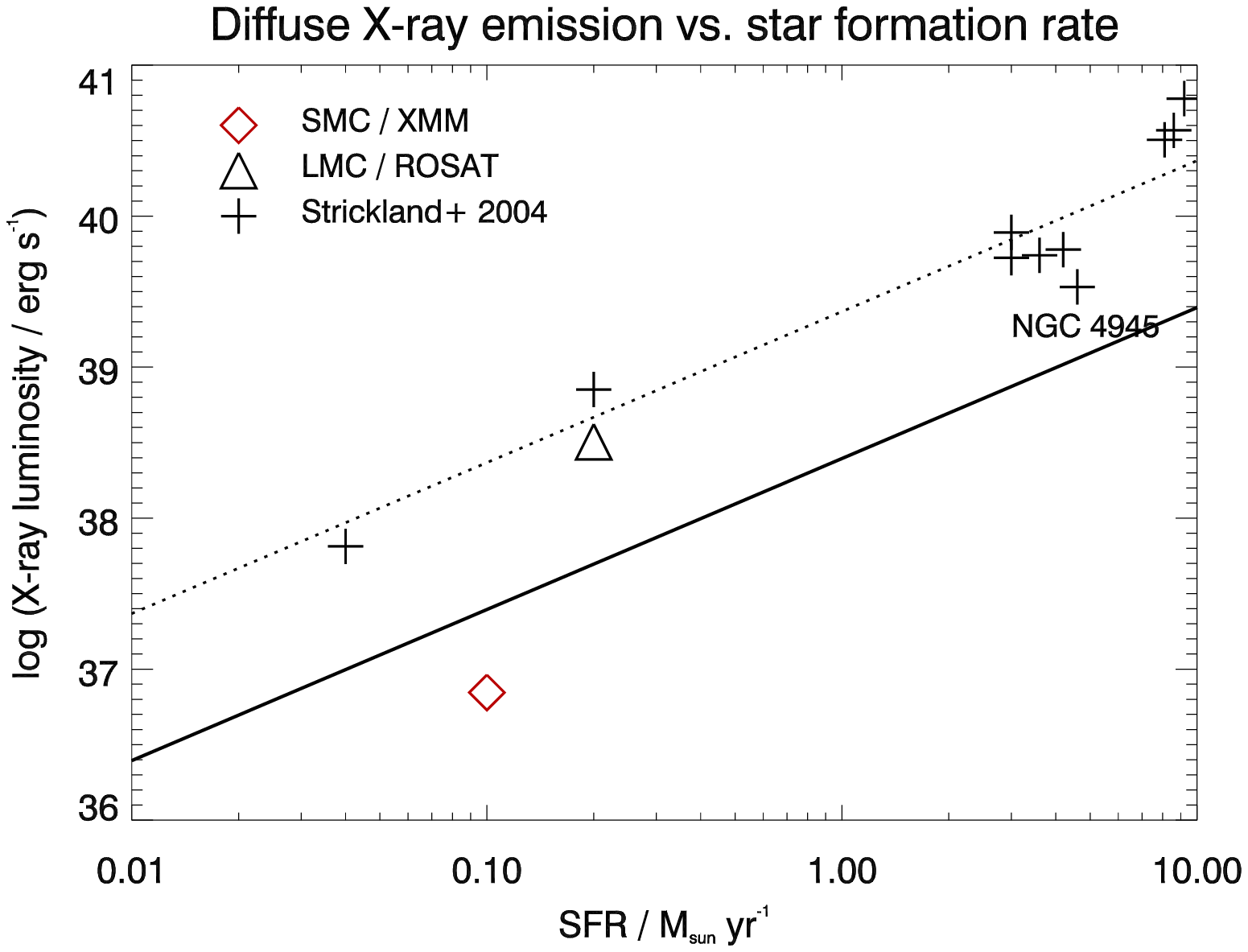}
      \caption{Diffuse X-ray luminosity in the 0.2-2 keV band
	versus star formation rate in nearby galaxies.
              Black plus signs are data from \citet{Strickea04a} for a sample of nearby galaxies
	with a \rv{large} range of star-formation properties. The black triangle shows the \rosat
	measurement of the LMC form \citet{Sasea02}. The result from \xmm survey 
	of the Small Magellanic cloud \citep{Sturm12,Sturmea13} is indicated by the red diamond.
	The dotted line is a fit to the Strickland et al. data. 
	The source which is furthest below this line (NGC~4945) is marked. 
	The solid line is the prediction from 
	our analysis for solar metallicity. It is below the fit line for the Strickland et al. 
	data by a factor of 9, which would decrease to 6 if the 0.1-0-2 keV band would be 
	added. For the SMC metallicity, 
	the prediction would be lower by a factor of ten.
	See sect.~\ref{sect:disc} for possible explanations for the discrepancy.}
         \label{fig:neargals}
   \end{figure}
%

We compare this to observations of nearby galaxies in Fig.~\ref{fig:neargals}.
The diffuse X-ray luminosity in these galaxies correlates with the star-formation rate,
as expected, if the X-ray emission is caused by a process related to star formation.
The \rosat value for the LMC is close to the fit line to the \chandra sample of
\citet{Strickea04a}, whereas the X-ray luminosity of the Small Magellanic Cloud 
(SMC) is a factor of 30 lower.
One reason for this is that the proximity of the SMC together with the 
spectral resolution capabilities of 
the \xmm telescope ensure the best possible background separation. 
These efforts have lead to a reduction of X-ray luminosity compared to 
the previous \rosat measurement of diffuse X-ray 
luminosity for the SMC \citep{Sasea02} by almost 40 per cent. It is possible 
that accounting for unresolved 
background sources would also lead to a reduction of the LMC luminosity. 
The same may even be true for the luminosities determined with \chandra for the sample of \citet{Strickea04a}, because the better spatial resolution of \chandra is compensated  by the larger distances. We note that the source which lies furthest below
the fit line, NGC~4945, has the best point source sensitivity within the \chandra sample,
$6\times 10^{36}$~erg~s$^{-1}$. 
Another reason why the SMC data point is so much below the correlation defined by the 
other data points is certainly the low metallicity of the ISM in the SMC of only 
$1/5$ of the solar value \citep{RD92}. This affects the total X-ray luminosity by
almost 1~dex \citep{SD93}. Our prediction would hence have to be adjusted about 
1~dex downwards, leading to a consistent underprediction of the 
soft X-ray luminosity in all these galaxies.
In the following, we adopt the viewpoint that the 
reported X-ray luminosities (apart from the SMC result) may perhaps be a factor of 
two to large, due to unaccounted backgrounds. 
In this case they would still be significantly above our predicted line.

%
\section{Discussion}\label{sect:disc}

We predicted the X-ray properties of superbubbles based on 3D hydrodynamics
simulations. The essential assumption
behind this approach is that the X-ray properties are a consequence of the 
repeated basic sequence:
shock-heating -- shell acceleration and hydrodynamic instabilities -- mixing.

Several observed properties are reproduced very well by our simulations:
\begin{itemize}
\item The X-ray luminosity of superbubbles is extremely variable in space and time. We find variations 
of two orders of magnitude in total X-ray luminosity. This agrees very well with 
the observation that some superbubbles are X-ray bright and some are X-ray dim
\citep[e.g.][]{Jaskea11}.
\item The peak X-ray luminosity is about $10^{36}$~erg/s. This a direct consequence
of the interplay of shock-heating and large-scale instabilities with subsequent mixing 
well modelled in our 3D simulations. This agrees well with 
observations (Fig.~\ref{fig:emwt}).
\item The predominant temperature of the X-ray emitting material is (for most of the time) sub-keV.
This difference compared to expectations from simple analytic bubble models
\citep[e.g.][]{DPC01} is due to the enhanced mixing due to Rayleigh-Taylor instabilities 
triggered by the strong dependence of the kinetic energy input on time.
\item Large-scale (in space) spectral changes are expected due to global oscillations
inside the bubble. We suggest this is the case in the nearby Orion-Eridanus superbubble.
\end{itemize}

\noindent
There are also disagreements with observations:
\begin{itemize}
\item The X-ray luminosity seems to decay to fast.  We infer this from the 
large number
of X-ray bright superbubbles, and also from the underpredicted
X-ray luminosities of nearby galaxies.
\item While we cover part of the region in the luminosity-temperature diagram 
(Fig.~\ref{fig:emwt}), the temperature of the X-ray emitting gas 
in observed superbubbles frequently seems to be a factor of a few higher than what we predict.
\end{itemize}

\noindent
How can we explain these disagreements?

We believe the disagreement can hardly be related to numerical issues: in the relevant 
phases, the derived properties are well converged, and it is clear that the fact that 
the X-ray luminosity is not fully converged in the very low luminosity phases does not
have any effect on the X-ray properties of any luminous phase later on 
(appendix~\ref{a:rescomp} and Fig.~\ref{fig:3S1rescomp}).

The ambient density might differ from the one we assumed (10~$m_p$~cm$^{-3}$).
This would change the density of entrained shell filaments and the growth rate of the
Rayleigh-Taylor instability, because the shell dynamics is linked to the ambient density.
We investigated this briefly with two additional simulations using one $25~M_\odot$ star,
for which we varied the ambient density by a factor of ten upwards and downwards compared
to our standard value. We found a factor of a few higher peak luminosity for the 
higher ambient density run. However, we believe this should be verified at finer resolution.
This issue is beyond the scope of the present work.
For N~154, \citet{Sasea11} derive an ambient density of 13~$m_p$~cm$^{-3}$,
quite close to the value we used. Still, they find a temperature and luminosity value
significantly outside the region covered by our simulations (Fig.~\ref{fig:emwt}).

Observed X-ray-bright superbubbles usually contain more massive stars than
the three we assumed. The result of our simulations is that 
the X-ray bright phases are essentially independent.
This should remain true also for richer star clusters, where the supernovae might
follow one another more rapidly as long as the X-ray bright phases associated with the
individual supernovae do not overlap. This condition should be satisfied if the time span between
consecutive explosions exceeds about 1~Myr, corresponding 
to about 30 massive stars powering a given superbubble. It is exceeded in many objects,
e.g. N~154 \citep{Sasea11}. Moving the luminosity peaks closer together 
(e.g. Fig~\ref{fig:conf}) may keep the individual superbubbles close to the peak luminosity,
but would hardly affect the cumulative energy emitted in X-rays, and thus the 
diffuse X-ray luminosities for nearby galaxies (Fig.~\ref{fig:neargals}) would still 
be underpredicted.

These findings might point to an agent which reduces mixing in observed 
superbubbles after a given shock-heating event: the main reasons for the sharp decline 
in X-ray luminosity after a maximum are adiabatic expansion and mixing with entrained material 
(radiation losses are negligible). Less mixing after the passage of the supernova shock
wave would therefore keep the luminosity
high for a longer period. This would of course also keep the temperature at a higher level, as 
required. It might also keep the surface brightness during the global oscillations higher,
as seems to be required for the case of the Orion Eridanus superbubble.

A magnetic field of significant strength might reduce the mixing via a suppression of instabilities
\citep[e.g.][]{JNS95}. This might happen on large scales, as well as on sub-resolution scales.
The magnetic field is likely strong in superbubbles: The expansion of the ejecta will
produce some magnetic energy via field line stretching. Subsequent turbulence
should then randomise the field components. These mechanisms successfully explain
the magnetic fields in the lobes of extragalactic radio sources \citep{Gaiblea09,HEKA11a}.
The ambient material is strongly compressed in the expanding shell. Since much of the
internal energy is lost due to radiation, the shell material might actually be supported 
by the magnetic field \citep{DS96a}. 
Magnetic fields in superbubbles have been observed \citep{Heald12}. Their effect has also been 
studied in simulations 
\citep[e.g.][and references therein]{Stilea09}, but implications for the 
\rv{thermal} X-ray properties have so
far not been investigated.
However, because mixing is important to produce the intermediate density regions which are 
responsible for the X-ray emission in the first place, magnetic fields may not entirely suppress the instabilities.

   \begin{figure}
   \centering
     \includegraphics[width=\hsize]{\fdir/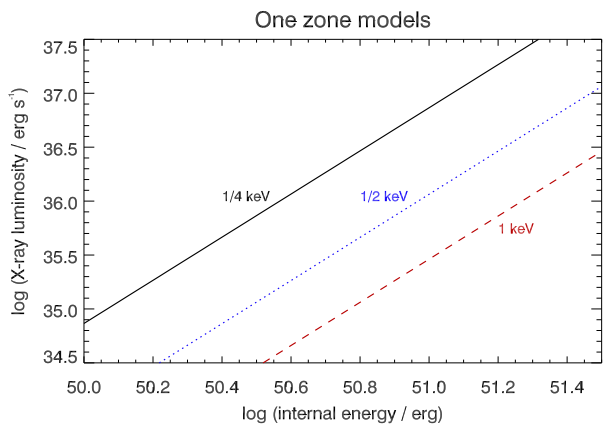}
      \caption{X-ray luminosity versus total internal energy for simple one-zone emission models.
	See sect.~\ref{sect:disc} for details.}
         \label{fig:1zone}
   \end{figure}
%

After a short X-ray bright phase, our simulated superbubbles dim in X-rays. At the same time,
they also loose their energy due to bubble dynamics: the expansion work on the
shell is then lost to (optical) radiation. The typical energy content is $10^{51}$~erg
in the X-ray bright phase and $10^{50}$~erg otherwise. The luminosity $L$ is related
to the heat energy content $E$ by $L = ((\gamma-1)E/kT)^2 \Lambda/V$, where $\Lambda$ denotes the cooling
function, $\gamma=5/3$ the adiabatic index and $V$ the volume. For an order of magnitude estimate, let us take $\Lambda$
from the solar metallicity models of \citet{SD93} in collisional ionisation equilibrium, 
and set the volume to $V= 4\pi (50\, \mathrm{pc})^3/3$. The resulting curves are shown in Fig.~\ref{fig:1zone} for 
kT~$=$ 1/4, \rv{1/2} and 1~keV. The plot demonstrates 
that with a bubble energy of $10^{51}$~erg, a mixture of 1/4-1~keV plasma can account 
for the observed X-ray luminosities of $10^{35}-10^{36}$~erg~s$^{-1}$. With $10^{50}$~erg,
this is not possible. This suggests that the dynamical energy loss time in real superbubbles 
is delayed, probably by a factor of a few, compared to our simulations.

An alternative explanation for the underprediction of the soft X-ray luminosity in galaxies
may of course be that other sources contribute also to the observed luminosity.
\citet{Persea04} have suggested that high-mass X-ray binaries may also contribute to the 
soft X-ray emission. We have noted earlier that point source contamination is indeed
an issue for galaxy-scale observations. However, so far the best available observations
are still above the luminosity we would predict (compare sect.~\ref{ssect:nearbygals}).

\rv{Some superbubbles have non-thermal contributions to the X-ray spectra from Cosmic rays
(compare Sect.~\ref{sec:intro}). Taking \object{30~Doradus} as example, one would expect
at most a similar non-thermal X-ray luminosity than the thermal one
\citep{Bambea04}. Because not all
superbubbles are detected in X-rays \citep{Yamea10}, it appears unlikely that
a non-thermal contribution could explain the discrepancy.}

The microscopic mixing process itself is not treated explicitly in our analysis. We simply assume here
that the gas phases are microscopically mixed at the resolution limit.
We believe this is justified, because the smaller scale instabilities responsible for the microscopic
mixing should be faster than the instabilities we resolve. Nevertheless, mixing of gas phases
will introduce a temporary non-equilibrium ionisation structure. This is expected to modify the
emissivity of certain emission lines. Effects are however expected mainly in the UV and EUV
part of the spectrum, and thus below the soft X-ray regime we consider here
\citep{BH87}.
   \begin{figure}
   \centering
     \includegraphics[width=.9\hsize]{\fdir/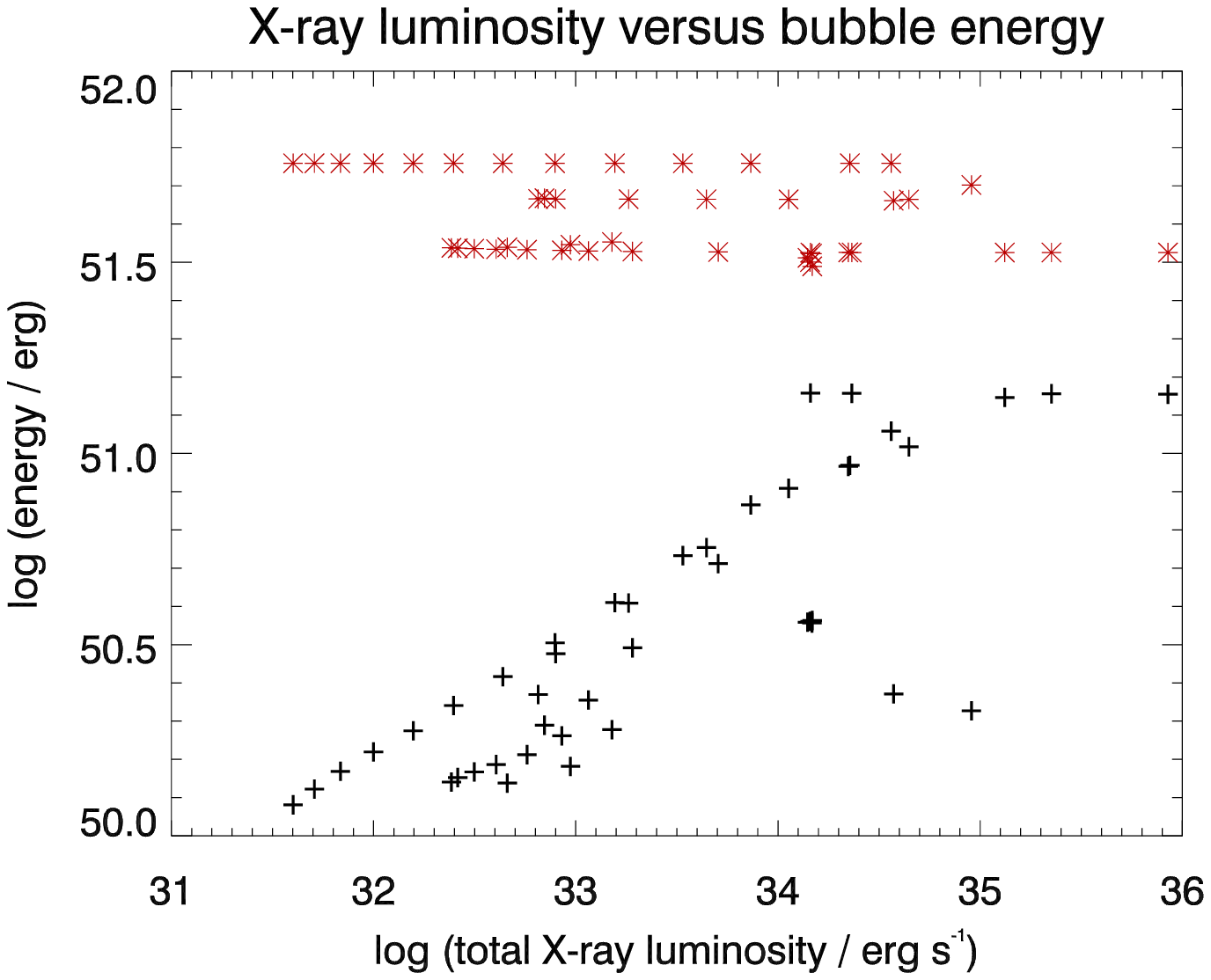}
     \includegraphics[width=.9\hsize]{\fdir/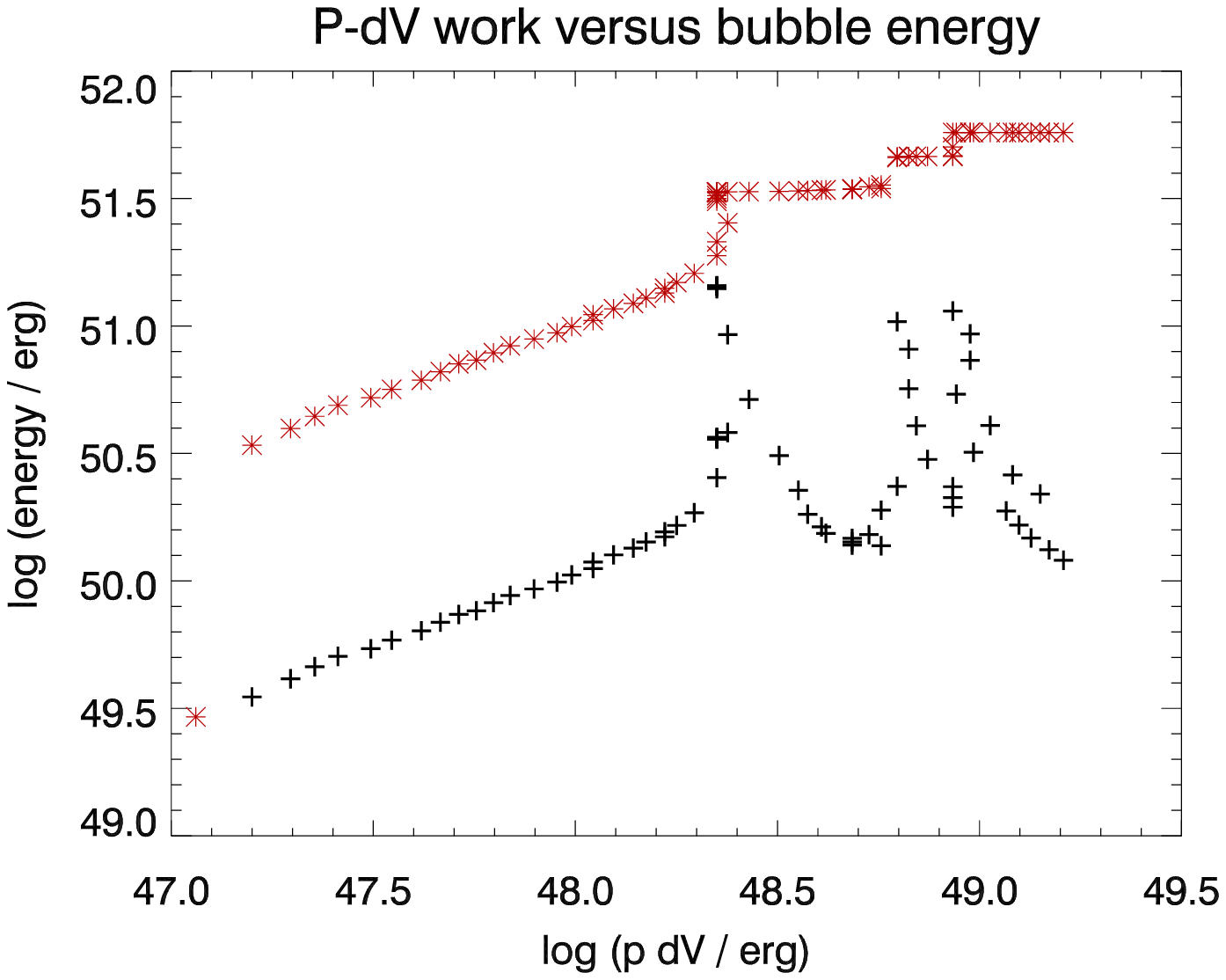}
     \includegraphics[width=.9\hsize]{\fdir/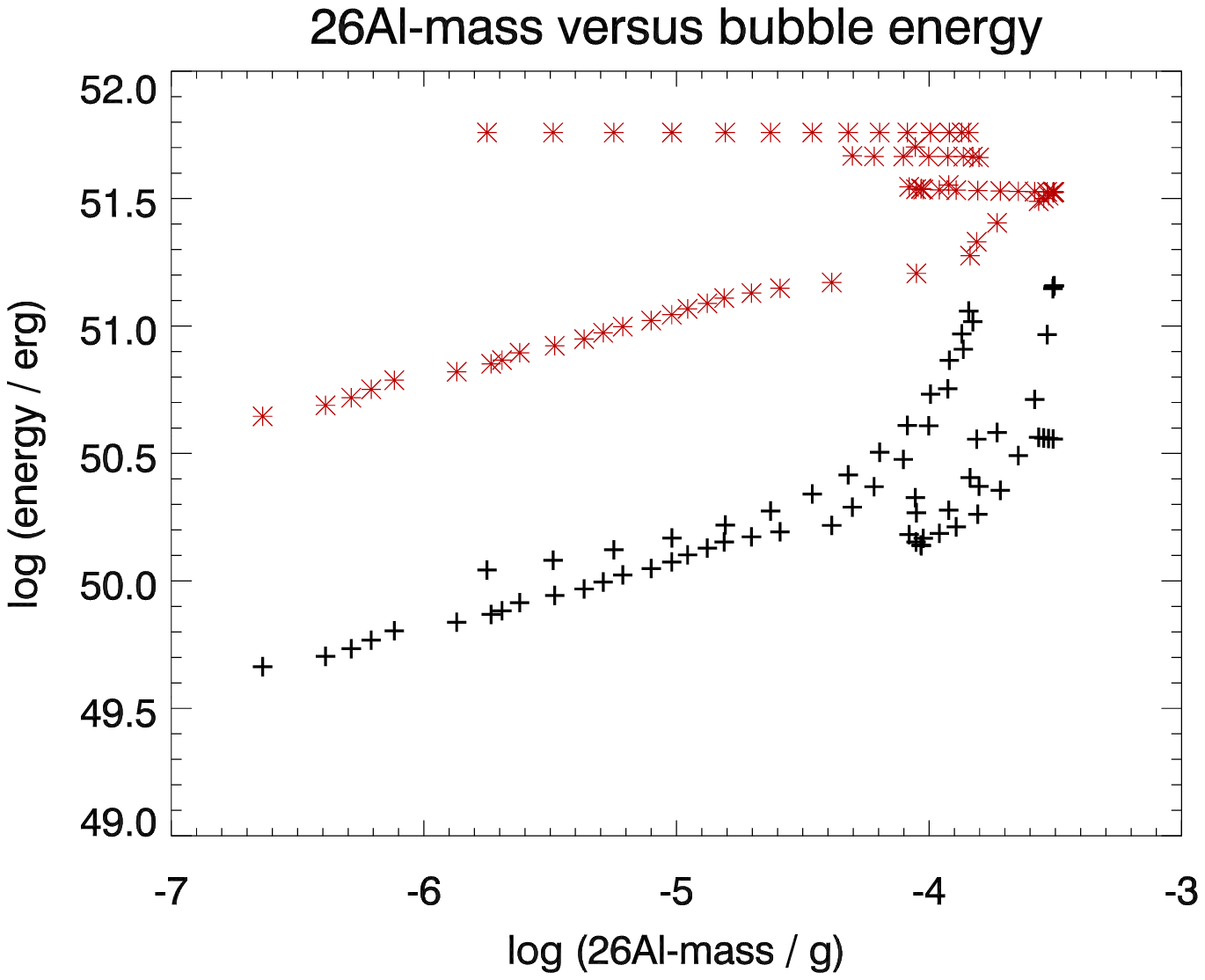}
      \caption{Scatter plots of the cumulative input energy (red stars) 
	and the total current energy (black pluses), both at a given time, 
	over a potential tracer on the horizontal axis for the representative simulation
	3S1-hr. For the latter, 
	we use the X-ray luminosity (top), the work against the ambient
	pressure (middle), and the mass of the radioactive isotope ${}^{26}$Al.
	See text for details.}
         \label{fig:etrac}
   \end{figure}
%

We have argued that high X-ray luminosities are a signature of large internal energy content of the superbubbles. It would
therefore be interesting to have a comparable observational tracer of the bubble energy.

In Fig.~\ref{fig:etrac}, we compare several possible tracers
with the cumulative input energy and the current energy in the simulation 
box, both at a given time.
As expected from our results, the X-ray luminosity does not correlate with 
the cumulative energy input.
For an individual superbubble, it makes therefore no sense to compare the 
X-ray luminosity to the stellar energy input (in contrast to the case for an 
ensemble of superbubbles when considering the 
X-ray luminosity of an entire galaxy, as discussed above).
However, the  X-ray luminosity does correlate with the current energy, present in a superbubble
at any given time, though with some scatter. The work against the ambient pressure may be estimated 
from HI-holes associated with the superbubbles \citep[e.g.][]{Bagea11}. Here, we defined the superbubble 
radius from the density peak in the column density map, similar to what one would do for HI observations.
The work done by the superbubble correlates very well with the cumulative input energy, but not with the 
current energy at a given time. Thus, HI-observations may inform us about the total energy released by the 
relevant group of massive star and provide us with complementary information with regard to X-rays.
Massive stars emit ${}^{26}$Al in their winds and supernovae. ${}^{26}$Al emits a 
Gamma-ray line 
\rv{which has been used to used to constrain the stellar content of superbubbles \citep[e.g.][]{Diehlea10, Vossea10,Vossea12}, infer the star-formation rate of the Milky Way \citep{Diehlea06}, and as kinematics tracer for hot, recently ejected gas \citep{Kretschea13}.} 
We used the data from stellar models as compiled 
in \citet{Vossea09} to relate the total signal expected from our simulated groups of stars to their mechanical
energy output \rv{as well as the retained energy at a given time.}(Fig.~\ref{fig:etrac}, bottom). 
\rv{As expected, the cumulative energy input may vary by more than an order of magnitude
for the same ${}^{26}$Al-mass. However, as ${}^{26}$Al is correlated to supernovae and strong
wind phases, there is a correlation with the current energy of the superbubble, though with some scatter, because the (exponential) radioactive decay differs from the (power law) radiative energy loss of the superbubble. Because of the correlation of the current bubble energy with the 
X-ray luminosity (compare above), one would essentially expect the ${}^{26}$Al-detected
superbubbles to be X-ray bright. This is indeed the case for the three individual superbubbles
where ${}^{26}$Al has been firmly detected, Cygnus \citep{Cashea80,Martinea10,Xuea13}, Orion \citep[and this paper]{Diehl02,Vossea10}, and Scorpius-Centaurus \citep{Snowdea97,Diehlea10}.}

Our simulations also allow an estimate for the optical emission-line luminosity
related to the shocks driven by the superbubble shells. 
Again, the reason why we can do this estimate is that the energy tracks after each supernova 
explosion are similar, and the stored energy in a superbubble does not accumulate 
with the number of explosions (compare also Paper~I).
In our simulations, 90 per
cent of the injected energy is radiated by the shell. Assuming that 7 per cent of this energy is emitted
in H$\alpha$, appropriate for slow shocks \citep{Innea87}, yields a superbubble-related
H$\alpha$-luminosity of about 
$L_\mathrm{SB}(\mathrm{H}\alpha)\approx 5\times 10^{40}$~SFR~erg~s$^{-1}$, 
where the star-formation
rate is taken in units of solar masses per year. This is not the dominant, but still a significant
fraction of the observed  luminosity 
$L_\mathrm{obs}(\mathrm{H}\alpha) =1.3\times 10^{41}$~SFR~erg~s$^{-1}$
\citep{Kenni98}.
Thus, while the shock-related part of the H$\alpha$-luminosity might plausibly trace the 
current energy content of a superbubble, the signal is likely strongly confused by effects of photoionisation.

\citet{Ackea11} use the Fermi satellite to identify a cocoon of cosmic rays in the nearby Cygnus superbubble.
Cosmic ray acceleration requires strong shocks \citep[e.g.][]{VY14}. The strong energy loss in our superbubbles
lead to comparatively low temperatures in the bubble interiors and thus increase the probability 
for newly formed supernova shocks to exceed the critical Mach number. However, since cosmic rays pervade the
Galaxy, we do not expect them to loose their energy similarly quickly than the thermal gas in the superbubble.
Thus, we would again expect that the Cosmic ray signal is more closely related to the total input energy than to the
current internal energy of a superbubble.

Therefore, it appears difficult to find another observational tracer with comparable properties to the soft X-ray luminosity\rv{, apart from ${}^{26}$Al which is, however, only detectable
close by in the \object{Milky Way}.}
Uniquely, \rv{soft X-rays trace} the current energy content, and not the cumulative mechanical energy.

%
\section{Conclusions}\label{sect:conc}

From the time-dependence of the energy injection into superbubbles
and the 3D nature of the hydrodynamics, we are able to reproduce the basic 
X-ray properties of superbubbles: strong variation in X-ray luminosities 
with space and time, peak luminosities
of the order of $10^{36}$~erg~s$^{-1}$, sub-keV temperatures and spatially varying
spectral properties, which we relate to global oscillations.

The analogy between the dynamics of superbubbles and the one of winds with constant
energy input rate \citep{Weavea77}, which has been made in the literature, is inadequate in several respects:
the X-ray luminosity is strongly linked to the time variability of the energy input rate; 
consequently, hydrodynamic 
instabilities and mixing are important. Including these effects, we reproduce the peak luminosities 
and soft spectra observed in superbubbles.
 
The X-ray emission from our simulated superbubbles, however, 
fades too quickly and has slightly too low temperatures.
This leads to an underprediction of the diffuse X-ray luminosity of nearby galaxies 
by about a factor of ten. We suspect that this may be due to suppression
of mixing after the shock-heating, possibly related to magnetic fields.
An alternative explanation would be additional contributions from other sources,
e.g. high-mass X-ray binaries \citep{Persea04}.


We find that the soft X-ray emission probably uniquely traces the current energy content of
superbubbles \rv{(except for ${}^{26}$Al for nearby objects)}, whereas other tracers correlate much better with the cumulative mechanical 
energy input from a group of massive stars.
 
\begin{acknowledgements}
  \rv{We thank the anonymous referee for useful comments.}
This research was supported by the cluster of excellence ``Origin
  and Structure of the Universe'' (www.universe-cluster.de). 
We acknowledge the use of NASA's {\it SkyView} facility
     (http://skyview.gsfc.nasa.gov) located at NASA Goddard
     Space Flight Center.
\end{acknowledgements}

\bibliographystyle{aa}
\bibliography{/Users/mkrause/texinput/references}

\appendix
\section{Effects of numerical resolution}\label{a:rescomp}

   \begin{figure}
   \centering
   \includegraphics[width=\hsize]{\fdir/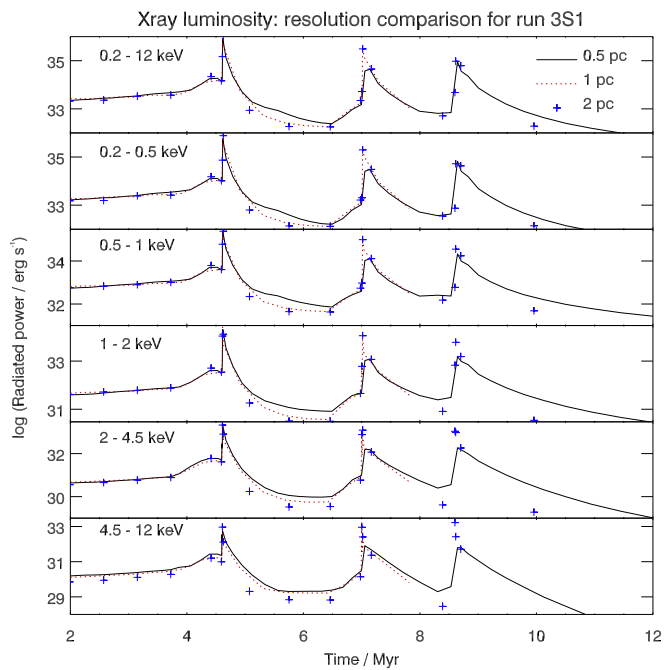}
       \caption{Integrated X-ray luminosity over time for different spatial resolutions and different energy bands indicated in the respective individual plots.
       The solid black line is for the highest resolution run, shown in 
       Fig.~\ref{fig:overview} (0.5~pc resolution for finest AMR level). 
       The dotted red curve is for a twice coarser grid, the blue pluses for again a factor of two worse resolution.              }
         \label{fig:3S1rescomp}
   \end{figure}
%

We repeated run 3S1-hr at lower resolution, by reducing the maximum 
refinement level by one and two, respectively. Figure~\ref{fig:3S1rescomp} 
shows the X-ray luminosity in various bands over time for the three simulations. 
Over most of the time, the simulation
has converged. Exceptions (for technical reasons) are the X-ray peaks (and minima), 
because in these short phases, the snapshot closest to the peak is offset in physical time 
by various amounts. Moderate differences occur in phases of low and declining luminosities.
However, even after these moderate discrepancies the luminosity returns to essentially 
the same level at each increase in energy input, i.e. whenever the bubble 
achieves high luminosities.

\end{document}